\documentclass[12pt]{article}
\usepackage[dvips]{graphicx}
\usepackage{amssymb, amsmath, mathbbol}
\usepackage[centerlast,footnotesize]{caption2}

\newcommand{\lapprox}{\raisebox{-0.5ex}{$\
\stackrel{\textstyle<}{\textstyle\sim}\ $}}
\newcommand{\gapprox}{\raisebox{-0.5ex}{$\
\stackrel{\textstyle>}{\textstyle\sim}\ $}}

\usepackage{graphicx}
\newcommand{\One}{1\kern-4.5pt1}

\newcommand{\be}{\begin{equation}}
\newcommand{\ee}{\end{equation}}

\def\lesim{${\lower 2pt\hbox{$\scriptstyle
<$}\atop\raise 4pt\hbox{$\scriptstyle\sim$}}$} 
\def\grsim{${\lower2pt\hbox{$\scriptstyle >$} \atop\raise4pt\hbox 
{$\scriptstyle\sim$}}$} 
\begin{document}
\begin{center}
\begin{flushright}
February 2013
\end{flushright}
\vskip 10mm
{\LARGE
Monte Carlo Study of Strongly-Interacting Degenerate Fermions:\\
a Model for Voltage-Biased Bilayer Graphene\\ 
}
\vskip 0.3 cm
{\bf Wes Armour$^{a}$, Simon Hands$^b$ and Costas  Strouthos$^c$}
\vskip 0.3 cm
%$^a${\em Diamond Light Source, Harwell Campus,\\
%Didcot, Oxfordshire OX11 0DE, United Kingdom.}
%\vskip 0.3 cm
$^a${\em 
Oxford e-Research Centre, University of Oxford, 7 Keble Road, Oxford OX1 3QG, United Kingdom.}
\vskip 0.3 cm
$^b${\em Department of Physics, College of Science, Swansea University,\\
Singleton Park, Swansea SA2 8PP, United Kingdom.}
\vskip 0.3 cm
$^c${\em Computation-based Science and Technology Research Center,\\
The Cyprus Institute, 1645 Nicosia, Cyprus.}
\vskip 0.3 cm
\end{center}

\noindent
{\bf Abstract:} 
We formulate a model of $N_f=4$ flavors of relativistic fermion in 2+1$d$ in
the presence of a chemical potential $\mu$ coupled to 
two flavor doublets 
with opposite sign, akin to isopsin chemical potential in QCD. This is argued to 
be an effective theory for low energy electronic excitations in bilayer
graphene, in which an applied  voltage between the layers ensures equal
populations of particles on one layer and holes on the other.  The model is
then reformulated on a spacetime lattice using staggered fermions, and in the
absence of a sign problem, simulated using an orthodox hybrid Monte Carlo
algorithm. With the coupling strength chosen to be close to a quantum critical
point believed to exist for $N_f<N_{fc}\approx4.8$,
it is found that there is a region below saturation where both the carrier density and a particle-hole
``excitonic'' condensate scale anomalously with increasing $\mu$, much more rapidly
that the corresponding quantities in free field theory, while the conventional
chiral condensate is strongly suppressed. The
corresponding ground state is speculated to be a strongly-correlated degenerate
fermion system, with a remnant Fermi surface distorted by a superfluid excitonic
condensate.
The model thus shows
qualitatively different behaviour to any model with $\mu\not=0$ previously
studied by lattice simulation.

%\noindent
%PACS: 11.10.Kk, 11.15.Ha, 71.10.Fd, 73.63Bd
                                                                                
\vspace{0.5cm}

\newpage
\section{Introduction}

The study of quantum systems containing a non-zero density of conserved charge
beyond perturbation theory remains a challenging but fascinating problem.
Estimating Euclidean Green functions via Monte Carlo importance sampling of the
discretised path integral is rendered inoperable once $\mu/T\gg1$, where $\mu$
is the chemical potential associated with the conserved charge such that the
charge density $n=\partial\ln{\cal Z}/\partial\mu$, because generically the
integrand $e^{-S_E(\mu)}$, with $S_E$ the action in Euclidean metric, is not
positive definite once $\mu\not=0$. There remain a small number of interesting
problems with real $S_E$ which can be tackled by orthodox methods: fermionic
models such as Gross-Neveu and NJL, which with $\mu\not=0$ are either Fermi
liquids~\cite{Hands:2003dh} or weakly-interacting BCS
superfluids~\cite{Hands:2001aq,Hands:2002mr}; certain gauge theories with
groups such as SU(2)~\cite{Cotter:2012mb} or G$_2$~\cite{Maas:2012wr} containing
real matter representations; and QCD with non zero isospin
density~\cite{Kogut:2002zg}. 
There is also a recent study of the interacting
relativistic Bose gas which deals with a complex $S_E$ by sampling a
complexified configuration space using Langevin dynamics~\cite{Aarts:2008wh}.

In the present paper we claim to add to this list a model of
strongly-interacting fermions in 2+1$d$ inspired by recent developments in
graphene. Recall that due to the special properties of the honeycomb lattice
(with just two isolated zeros or ``Dirac points''
of the dispersion $E(\vec k)$ in the first Brillouin
zone),
low energy excitations of either electron or hole nature are naturally described
by a 2+1$d$ Dirac equation. For monolayer graphene, the counting of degrees of
freedom (2 Dirac points $\times$ 2 C atoms per unit cell $\times$ 2 electron spin
components) results in $N_f=2$ flavors of 4-component relativistic spinor fields.
Here we consider a model of bilayer graphene comprising $N_f=4$
relativistic flavors, in which a biassing voltage is applied across the layers
so that a non-zero density of electrons is induced on the negative sheet and an
equal density of holes induced on the positive. It will be shown in
Sec.~\ref{sec:form} that this is equivalent to introducing an equal and opposite
chemical potential $\mu$ on each sheet, so that the resulting relativistic
effective theory has an ``isospin'' chemical potential $\mu_I$ with two ``$u$''
and two ``$d$'' flavors. Just as for QCD, it is
straightforward to show the resulting $S_E$ is real and positive so that Monte
Carlo simulation is applicable. 

The role for numerical simulation becomes apparent once inter-electron interactions
are considered. The Coulomb interaction between charges in graphene is effectively enhanced
by a factor $v_F/c\approx300$, where $v_F=\partial E/\partial k\vert_{k=k_F}$ is the Fermi
velocity of the charge carriers, which is constant for a linear dispersion. Since
in undoped graphene with carrier density $n_c=0$ 
only quantum fluctuation (namely vacuum polarisation) effects contribute
to screening, the resulting effective theory of relativistic
electrons has an effective fine structure constant $\alpha_{\rm eff}\sim O(1)$,
albeit with a non-covariant ``instantaneous'' interaction of the form
$A_0\bar\psi\gamma_0\psi$ due to $v_F\ll c$. This implies that the theory must
be treated as strongly-interacting, and the possibility of non-perturbative effects such as
particle-hole condensation $\langle\bar\psi\psi\rangle\not=0$, resembling chiral
symmetry breaking and resulting in dynamical mass gap generation leading to 
a Mott insulator phase, should be
considered. Son~\cite{Son:2007ja}, in the context of a model with $N_f$
relativistic species and variable coupling strength $\alpha$  has suggested that 
gap generation occurs for $N_f$ sufficiently small and $\alpha$
suffciently large. Moreover, the critical coupling $\alpha_c(N_f)$ defines a
quantum critical point (QCP) where the scaling of operators and correlation
functions in principle receive anomalous corrections, which could for instance modify the
linearity of the electron dispersion.

The scenario was tested in 
a Monte Carlo simulation with 2+1$d$ staggered fermions (at a QCP the
underlying lattice should become irrelevant)~\cite{Hands:2008id}; 
in the strong coupling limit $\alpha\to\infty$ the QCP is found for
$N_{fc}=4.8(2)$. Simulations incorporating a more realistic long-ranged Coulomb
interaction estimated $\alpha_c=1.11(6)$~\cite{D&L} for the case $N_f=2$
relevant for monolayer graphene, suggesting there is a real possibility of gap
generation for suspended samples (the effective value of $\alpha$ is sensitive
to the dielectric properties of the substrate, and is expected to be maximal
when the substrate is absent). To date there is no definitive experimental
signal for chiral symmetry breaking or mass gap generation; however,
non-linearities in the electron dispersion as a result of interactions have been
reported in \cite{Elias:2011}. Even if gap generation does not take place, it is
possible that 
physical graphene lies close to the QCP in the chirally symmetric phase,
implying modifications to 
electron transport. 

For the case of a voltage-biased bilayer a new condensation channel, between
particles in one layer and holes in the other, opens up; in what follows
we will refer to this
as ``excitonic'' condensation. Because of the
increasing density of states, as the Fermi energy is raised this should become
the preferred channel even though inter-layer interactions are weaker than
intra-layer ones. Once again, gap formation results implying an insulating
phase; since this can be controlled by the external voltage the possibility of 
graphene-based  electronic components arises~\cite{Castro:2010}. A
self-consistent treatment however, taking into account the enhanced screening due to
$N_f=4$, finds the resulting gap $\Delta/\mu\sim O(10^{-7})$, suggesting that
excitonic condensation will be difficult to achieve
experimentally~\cite{Kharitonov}.

Our purpose in this paper is to re-examine excitonic condensation in the
non-perturbative setting required by the vicinity of a QCP. We will reformulate the
lattice model used for the investigation of the QCP in undoped graphene in
\cite{Hands:2008id, AHS1} to include $N_f=4$ continuum Dirac fermions and a
nonzero isospin chemical potential $\mu$ to represent the voltage bias. In order
to formulate the model with a local interaction on a 2+1$d$
lattice, the
interaction between fermions is forced to have the form
of a contact between local charge densities, schematically 
$(\bar\psi\gamma_0\psi)^2$, and hence no long-range Coulombic tail. As argued
in \cite{Hands:2008id}, we expect that large vacuum polarisation effects make
this irrelevant near the QCP (or equivalently in the large $N_f$ limit). The
model
is presented in detail in Sec.~\ref{sec:form}. An important and necessary
simplification is that inter- and intra-layer interactions between
fermions have a
common coupling and hence the 
same stength. Since $\mu\not=0$ boosts the density of available particle-hole states, 
we again expect
this to become less relevant as $\mu$ is increased.

In Sec.~\ref{sec:results} we present results from numerical simulation of the
model on lattice volumes $32^3$ and $48^3$. Since it supposedly describes
a continuum effective theory in the vicinity of a QCP it is diffcult at this
exploratory stage to assign physical units to the simulation results, or even in
principle to
know the physical anisotropy ratio $a_t/a_s$. We choose a coupling strength
close to the putative QCP for $N_f=4$; since the strong-coupling limit is hard
to isolate for our formulation~\cite{Christofi:2007ye} this proves to be a
non-trivial exercise, and indeed we will see it appears our simulations lie just in
the chirally symmetric phase at $\mu=0$. Next, we introduce $\mu\not=0$ and
monitor the chiral condensate, carrier density and exciton condensate as it is
increased. It will be shown that excitonic condensation does indeed take place
and that both carrier density and exciton condensate considerably exceed
the values expected for free fermions (augmented by a small symmetry-breaking source
term), whereas the chiral condensate is suppressed.
We discuss our findings in Sec.~\ref{sec:discussion}. Although the model is
motivated by condensed matter physics, we will argue the results are of wider
interest, and yield perhaps the first insight into Fermi surface physics in the
presence of genuinely strong interactions.

\section{Formulation and Interpretation of the Model}
\label{sec:form}

Here we outline the formulation of an effective field theory for the graphene bilayer. 
Physically, the idea is that there are $N_f=2$ flavors of relativistic fermion on each
monolayer, described by an action in Euclidean
metric~\cite{Khveshchenko1,Son:2007ja}:
\begin{equation}
S_{\rm mono}=
\sum_{a=1,2}\int dx_0d^2x
(\bar\psi_a\gamma_0\partial_0\psi_a+v_F\bar\psi_a\vec\gamma.\vec\nabla\psi_a+iA_0\bar\psi_a\gamma_0\psi_a)+{1\over{2e^2}}\int
dx_0d^3x(\partial_iA_0)^2, 
\label{eq:Scont}
\end{equation}
where $e$ is the effective electron charge (whose value depends on the
dielectric properties of the substrate), 
and the $4\times4$ Dirac matrices satisfy
$\{\gamma_\mu,\gamma_\nu\}=2\delta_{\mu\nu}$, $\mu=0,1,2$. Note that for this 
reducible represention of the Dirac algrebra there exist two independent
matrices $\gamma_3$ and $\gamma_5\equiv\gamma_0\gamma_1\gamma_2\gamma_3$ which
anticommute with the $\gamma$-matrices present in (\ref{eq:Scont}).
$A_0$ is a
fluctuating $3+1d$ electrostatic potential field sourced by the charge density
$\bar\psi\gamma_0\psi$, and is a remnant of the full electromagnetic field in
the instantaneous approximation justified for $v_F\ll c$.

Now, for a perfect bilayer formed from two monolayers stacked in AB
configuration with interlayer coupling strength $t^\prime\sim O(0.1)t$ where
$t$ is the hopping parameter in the monolayer tight-binding Hamiltonian, it is
known 
that the dispersion relation for massless fermions in the low-energy limit in
the vicinity of the Dirac point is quadratic, only 
becoming approximately 
relativistic (i.e. linear) for $ka\gapprox t^\prime/t$~\cite{McCann}. For
$\mu\gg t$ the dispersion then takes the expected form $\varepsilon^2=(\mu\pm
v_Fk)^2$~\cite{CN2}.
However, recent theoretical studies of strained bilayers suggest
that under mechanical deformation the parabolic bands split to form separate
Dirac cones, so that in this case a description in terms of $N_f=4$ relativistic species is not
a bad approximation~\cite{strain}. Our formulation makes the additional, perhaps
unwarranted, approximation
that interactions between charge carriers on different layers are of identical strength and
character to interactions within a layer - the necessity for this will become
clear below.

The second ingredient of the model is that the layers are given equal and
opposite constant bias
voltages $\pm\mu$, inducing on one layer a negatively charged
concentration of particles and on the other a positively-charged concentration
of holes.  
As the notation implies, the bias voltage is equivalent to a chemical
potential, and in fact the theory is formally very similar to the case of QCD
with isospin chemical potential $\mu_I=\mu_1=-\mu_2$, where
the subscripts which here label the layers usually stand for the light quark
flavors $u$ and $d$. Euclidean formulations of systems with $\mu\not=0$ are
generally afflicted with a ``Sign Problem'', ie. the Lagrangian density ${\cal L}$
is no longer positive definite, or even real, since the inequivalence under
time reversal translates into inequivalence under complex conjugation in
Euclidean metric.  This makes Monte Carlo importance sampling as a means to
handle strongly fluctuating observables inoperable. However, the case of 
isospin chemical potential 
is known not to have a Sign Problem and is hence simulable using orthodox
methods, as we shall now demonstrate. 

If we denote the fermion degrees of freedom on one layer by $\psi$ and on the other by
$\phi$, define units so that $v_F=1$, and write
$\sum_{\mu=0,1,2}\partial_\mu\gamma_\mu+(iA_0+\mu)\gamma_0=D[A;\mu]$,
then the fermion part of the Lagrangian can be written: 
\begin{equation}
{\cal L}
= (\bar\psi,\bar\phi)\left(
\begin{matrix}D[A;\mu]+m&ij\cr
-ij&D[A;-\mu]-m\cr
\end{matrix}\right)\left(
\begin{matrix}\psi\cr\phi\cr\end{matrix}\right)\equiv\bar\Psi{\cal M}\Psi
\label{eq:action}
\end{equation}
Here we have introduced two new real parameters:
$m$ is an artificial bare mass which induces a gap in the 
fermion dispersion relations, and whose sign has no physical consequence for a single
flavor
in the absence of interactions; $j$ a
source strength coupling $\psi$ to $\phi$, thus linking the layers
and eventually
enabling calculation of the exciton condensate. In principle both $m\to0$ and
$j\to0$ limits need to be taken in order to make contact with physical bilayer
graphene. Integration over the Grassmann bispinors $\Psi,\bar\Psi$
then results in the functional measure $\mbox{det}{\cal M}[A]$.

An important identity which the model inherits from the gauge theory is
\begin{equation}
D^\dagger[A;\mu]=-D[A;-\mu].
\label{eq:conjg}
\end{equation}
It is then straightforward to check (assuming the dimension of $D$ is even) that
\begin{equation}
\mbox{det}{\cal M}=\mbox{det}[(D+m)(D+m)^\dagger+j^2]>0,
\end{equation}
and
\begin{equation}
{\cal M}^\dagger{\cal M}=\left(\begin{matrix}
(D+m)^\dagger(D+m)+j^2&\cr
&(D+m)(D+m)^\dagger+j^2\cr\end{matrix}\right),
\end{equation}
implying both that 
\begin{equation}
\mbox{det}{\cal M}^\dagger{\cal M}\equiv\mbox{det}^2{\cal M},
\end{equation}
and also that the desired functional measure $\mbox{det}{\cal M}$ results from integrating over
bosonic fields $\Phi$ starting from a non-local ``pseudofermion''
Lagrangian
\begin{equation}
{\cal L}_{pf}=\Phi^\dagger[(D+m)^\dagger(D+m)+j^2]^{-1}\Phi.
\label{eq:sim}
\end{equation}
This has the practical advantage that $\Phi$ has half as many degrees of freedom
as $\Psi$, and makes eqn.~(\ref{eq:sim}) the appropriate starting point for the
design of a hybrid Monte Carlo simulation algorithm.

The specific version of $D+m$ in our lattice model employs single-component
staggered fermion
fields $\psi_x$, $\phi_x$ defined on the sites of a $2+1d$ square lattice, with
a {\it non-compact} formulation of the electrostatic potential $A_x$
formally defined on the link joining sites $x$ and $x+\hat0$:
\begin{equation}
(D+m)_{xy}=\sum_{i=1,2}{\eta_{i x}\over2}[\delta_{y,x+\hat\imath}
-\delta_{y,x-\hat\imath}]+{\eta_{0x}\over2}[(1+iA_x)e^\mu\delta_{y,x+\hat0}-(1-iA_{x-\hat0})e^{-\mu}
\delta_{y,x-\hat0}]+m\delta_{xy},
\label{eq:slatt}
\end{equation}
where the signs $\eta_{\mu x}=(-1)^{x_0+\cdots+x_{\mu-1}}$ ensure Lorentz covariance in
the long wavelength limit. It can be shown that the relation between the number
of staggered fields $N$ (counting $\psi$,$\phi$ yields $N=2$) and the number $N_f$ of continuum Dirac 4-spinors
is~\cite{BB}
\begin{equation}
N_f=2N.
\end{equation}

It is worth noting the global symmetries present in the model. For $\mu=m=j=0$
the continuum action (\ref{eq:action}) is invariant under a U(8) rotation
$\Psi\mapsto U\Psi$, $\tilde\Psi\mapsto\tilde\Psi U^\dagger$ where
$\tilde\Psi\equiv i\bar\Psi\gamma_3\gamma_5$. This symmetry is broken to
(U(4))$^2$ by $\mu\not=0$, and then to (U(2))$^4$ by $m\not=0$. Setting the
interlayer coupling $j\not=0$ with $m=0$ locks the $\psi$ and $\phi$ components together,
so that in this case the residual symmetry is U(4). For the staggered
lattice fermions of (\ref{eq:slatt}) the original symmetry is
U(2)$\otimes$U(2)$_\varepsilon$, where the second rotation can be written as
$U(\alpha;x)=\exp(i\varepsilon_x\alpha^a\Lambda^a)$, where $\Lambda^a$ is one of the four
hermitian generators of U(2) and $\varepsilon_x\equiv(-1)^{x_0+x_1+x_2}$.
Setting $\mu\not=0$ breaks this to (U(1)$\otimes$U(1)$_\varepsilon$)$^2$,
followed by $m\not=0,j=0$ to (U(1))$^2$, and $m=0,j\not=0$ to
U(1)$\otimes$U(1)$_\varepsilon$.

The fermion action is supplemented by a Gaussian weight for the $A$ fields:
\begin{equation}
S_{\rm aux}={N\over4 g^2}\sum_x A_x^2,
\label{eq:Sbos}
\end{equation}
where $g^2$ is a parameter governing the strength of the coupling between the
potential and the fermions. The resulting dynamics describes $A$ 
fluctuations having the same form as the continuum action (\ref{eq:Scont})
in the strong-coupling or large-$N_f$ limits, but for which explicit screening
removes the long-ranged $r^{-1}$ tail away from these limits; further  justification
for this approximation is given in \cite{Hands:2008id,AHS1}.
For $N_f=2$ this formulation yields an identical path integral to the lattice
action couched in terms of compact link variables given in Eqn. (7) of
\cite{AHS1}.  For $N_f>2$, however, the two approaches are not equivalent since
the compact formulation leads to extra terms of the form
$(\bar\psi\psi\bar\phi\phi)^2$ in the effective action -- although these
operators may well be irrelevant at the critical point.  The exact lattice
version of the non-compact action for $\mu=0$ and arbitrary $N_f$ once $A$ is
integrated out is given in eqn.  (2.2) of \cite{LDD1}.

We next discuss the implications of relaxing the requirement that inter- and
intralayer interactions between fermions are identical.
The non-trival terms in the action are of the form $\bar\psi Ue^\mu\psi$,
$\bar\psi U^*e^{-\mu}\psi$, where $U$ is a complex number {\it not}
constrained to have unit modulus. Integration over $U$ leads to repulsive
particle-particle and hole-hole interactions, and attractive particle-hole
interactions.
Suppose we wanted to make the model more realistic by introducing a distinction
between intralayer and interlayer interactions. One way to do this would be to
introduce a second boson field coupling to $\psi$ and $\phi$ with
opposite signs, in effect introducing repulsion between $\psi$-particles and
$\phi$-holes so that the $\bar\psi$-$\phi$ and $\bar\phi$-$\psi$ couplings are 
weaker than those of
$\bar\psi$-$\psi$ or $\bar\phi$-$\phi$. The interaction terms could then be
written $\bar\psi_xUVe^\mu\psi_{x+\hat0}$, $\bar\phi_x UV^*e^{-\mu}\phi_{x+\hat0}$, 
$-\bar\psi_x
U^*V^*e^{-\mu}\psi_{x-\hat0}$, $-\bar\phi_x U^*Ve^{\mu}\phi_{x-\hat0}$, etc.
In the limit $V\to1$ integration over $\psi,\bar\psi$ leads to a factor
$\mbox{det}D[\mu]$, while integration over $\phi,\bar\phi$ gives
$\mbox{det}D[-\mu]$. With the help of (\ref{eq:conjg}) we confirm 
the resulting functional measure 
$\mbox{det}D[\mu]D^\dagger[\mu]$ is positive definite. In the limit
$U\to1$, however, the same process leads to $\mbox{det}D[\mu]D^*[-\mu]=
\mbox{det}^2D[\mu]$, which is no longer positive definite. In other words,
attempting to make the model more realistic reintroduces a Sign Problem,
although a more detailed study would be needed to determine its severity.

\begin{figure}[htb]
\begin{center}
    \includegraphics[width=10.0cm]{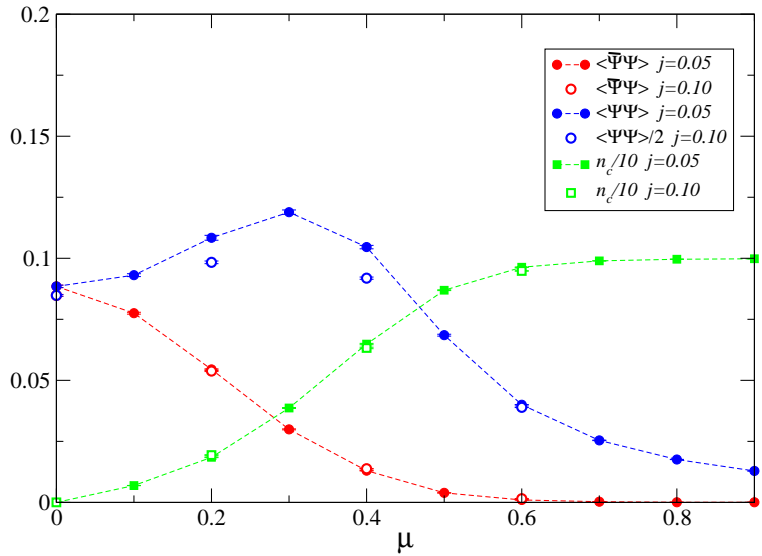}
\smallskip
\caption{Fermion condensates as a function of $\mu$
at $g^{-2}=0.4$ on $8^3$ with bare mass $ma=0.05$, $ja=0.05,0.1$.}
\label{fig:overview}
\end{center}
\end{figure}

Now let's discuss observables. The usual chiral condensate (which has been  
called the ``exciton condensate'' in our earlier work \cite{Hands:2008id,AHS1}) is given by
\begin{equation}
\langle\bar\Psi\Psi\rangle\equiv
{{\partial\ln{\cal Z}}\over{\partial
m}}=\langle\bar\psi\psi\rangle-\langle\bar\phi\phi\rangle.
\label{eq:chiral}
\end{equation}
Note the sign of the condensate is not physical, and that the two terms
on the RHS of (\ref{eq:chiral}) give equal contributions.
From the discussion above it should be clear that for $\mu\not=0$ formation of
this condensate spontaneously breaks (U(1)$\otimes$U(1)$_\varepsilon$)$^2$ to
(U(1))$^2$, resulting in two Goldstone modes in the limit $m\to0$, $j\to0$.
The exciton condensate discussed in
\cite{
Kharitonov} and which is the main focus of this paper
is given by
\begin{equation}
\langle\Psi\Psi\rangle\equiv{{\partial\ln{\cal Z}}\over{\partial
j}}=i\langle\bar\psi\phi-\bar\phi\psi\rangle.
\end{equation}
In this case the symmetry breaks to U(1)$\otimes$U(1)$_\varepsilon$
implying the same number of Goldstones.
In fact for $\mu=0$ and $m=j$, $\langle\bar\Psi\Psi\rangle$ and
$\langle\Psi\Psi\rangle$ are physically indistinguishable, both being equivalent
to the chiral condensate of the $N_f=2$ theory. Figure~\ref{fig:overview} 
below confirms that
with $\mu=0$ our code generates results consistent with
$\langle\bar\Psi\Psi\rangle/\langle\Psi\Psi\rangle\equiv{m\over j}$.

With $\mu\not=0$ we next define the charge carrier density
\begin{equation}
n_c\equiv{{\partial\ln{\cal Z}}\over{\partial\mu}}=\langle\bar\psi
D_0\psi\rangle - \langle\bar\phi D_0\phi\rangle.
\label{eq:cdensity}
\end{equation}
Once again, both terms on the RHS give equal contributions -- the first term
represents the density of electrons in layer 1, and the second the density of
holes in layer 2. 

Figure~\ref{fig:overview} shows the results of a pilot run on $8^3$ at
$g^{-2}=0.4$ and $ma=0.05$.
For $ja=0.05$ the two condensates are
degenerate at $\mu=0$ as argued above. As $\mu$ increases, our naive expectation
is that a
Fermi surface of radius $\mu$ forms on each layer, one containing particles,
the other holes, implying $n_c\propto\mu^2$. As $\mu$ grows, $\psi\bar\psi$ and
$\phi\bar\phi$ pairing are suppressed because a free particle-hole pair costs energy $2\mu$ to
create at either Fermi surface, whereas $\psi\bar\phi$ pairing is promoted, because it
costs zero energy to create a particle on one Fermi surface and a hole at the
other, with the
density of states at either increasing $\propto\mu$. Thus
$\langle\bar\Psi\Psi\rangle$ decreases as $\mu$ rises from 0, while
$\langle\Psi\Psi\rangle$ increases. The rise in $\langle\Psi\Psi\rangle$ seems
to be relatively more pronounced for smaller $j$.
This trend persists until $\mu a_t\simeq0.3$. What happens after that should
be understood in terms of {\it saturation}, an artifact which sets in
when the fermion density is a significant fraction of its maximal value of one
per lattice site. With our normalisation of $n_c$ 
this sets in for $\mu a_t\simeq0.5$, a surprisingly small value
based on experience with other models. In a saturated world fermion
excitations of all kinds are kinematically suppressed, and the condensates tend
to zero in this limit.

\section{Numerical Results}
\label{sec:results}

Our strategy in this paper is to investigate the effect of varying $\mu$ in our
bilayer model (\ref{eq:slatt},\ref{eq:Sbos}) 
starting close to the quantum critical point. The first task is
to find the coupling $g_c^2$ where the QCP is located for $N_f=4$; we use
the approach \cite{Christofi:2007ye,Hands:2008id} of searching for a maximum of
$\langle\bar\Psi\Psi\rangle$ as $g^{-2}$ is varied and identifying that with the
strong coupling limit of the continuum model. We then assume $g_c^{-2}\gapprox
g_{\rm peak}^{-2}$, since 
if the value $N_{fc}=4.8(2)$ obtained
in \cite{Hands:2008id} is universal there should only be a narrow
range of $g^{-2}$ corresponding to the chirally broken phase.
\begin{figure}[htb]
    \centering
    \includegraphics[width=10.0cm]{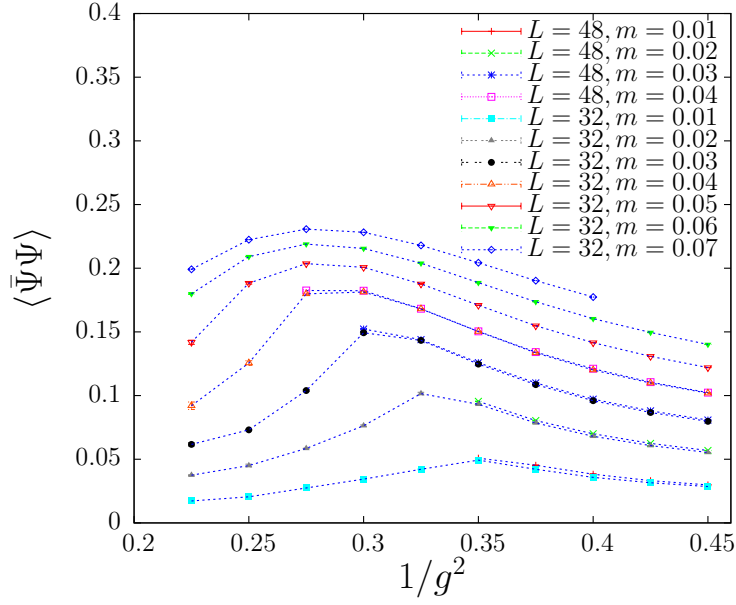}
    %\vspace{-6mm}
    \caption{(Color online)
$\langle  \bar{\Psi}\Psi  \rangle$ vs $g^{-2}$ for $N_f=4$ and various $m$ 
near $g^{-2}_{\rm peak} \approx 0.30$.
The simulations were performed on both $32^3$ and  $48^3$ lattices.}
\label{fig:gcrit}
\end{figure}
\begin{figure}[!ht]
    \centering
    \includegraphics[width=10.0cm]{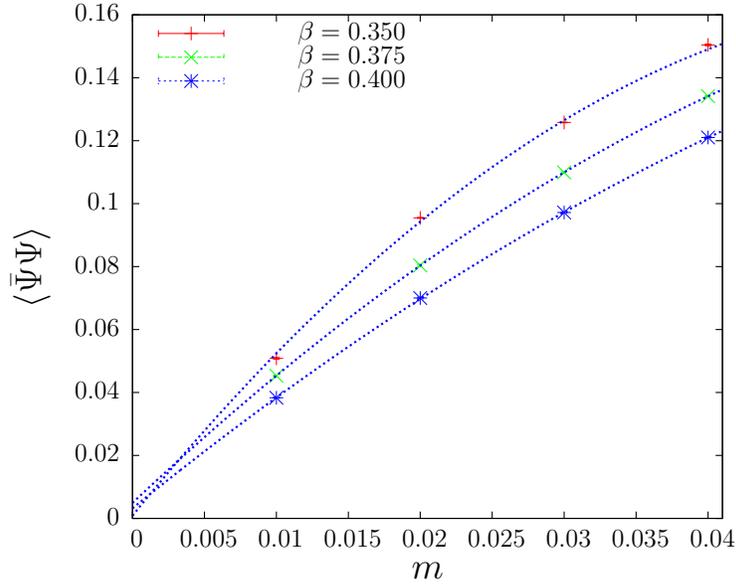}
    %\vspace{-6mm}
    \caption{(Color online)
$\langle\bar{\Psi}\Psi\rangle$ vs $m$ for $g^{-2}=0.35, 0.375, 0.40$ fitted to a 
quadratic polynomial. }
\label{fig:chirallimit}
\end{figure}
The results for $\langle\bar\Psi\Psi(m)\rangle$ in 
Fig.~\ref{fig:gcrit} show that $g_{\rm peak}^{-2}\approx0.30$, 
much larger than the value $\approx0.05$ obtained with the compact
formulation~\cite{Hands:2008id}. Another contrast with previous work is
that it is also apparent that 
$g^{-2}_{\rm peak}$ increases with $m$, from roughly $0.275$ at $ma=0.07$ to
0.35 for $ma=0.01$, although at this stage we cannot exclude the possibility that
finite volume effects influence the result.
For small $m$ a linear extrapolation to the chiral limit seems reasonable; we
conclude, conservatively, that in this limit $g^{-2}_{\rm peak}\in(0.275,0.35)$.

Figure~\ref{fig:chirallimit} shows
$\langle\bar\Psi\Psi\rangle$ data as a function of $m$ for $g^{-2}\approx
g^{-2}_{\rm peak}$. Whilst the quadratic extrapolation to the chiral limit is
not conclusive, the marked non-linearity of the fits suggests
the QCP value $g_c^{-2}$ lies close to this
region;  however, a much more extensive simulation campaign would be needed to pin it
down. For our purposes it suffices to work close to a strongly-interacting QCP,
while leaving the issue of whether chiral symmetry spontaneously breaks
unresolved.
Henceforth, all numerical results are obtained  with the coupling value
$g^{-2}=0.4$ -- this implies that the lattice cutoff is constant as $\mu$ is
varied. Unless otherwise stated, the chiral limit $m=0$ will be assumed.

\begin{figure}[htb]
    \centering
    \includegraphics[width=10.0cm]{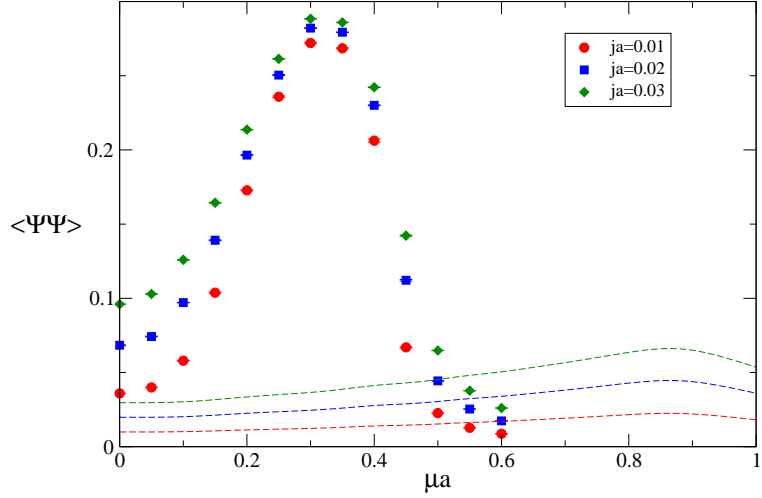}
    %\vspace{-6mm}
    \caption{(Color online)
$\langle \Psi\Psi\rangle$ vs $\mu$  on $32^3$ for $m=0$ and $ja=0.01, 0.02, 0.03$.
Dashed lines show the same
quantity evaluated for free fields.}
\label{fig:muscan_exciton}
\end{figure}
\begin{figure}[!htb]
    \centering
    \includegraphics[width=10.0cm]{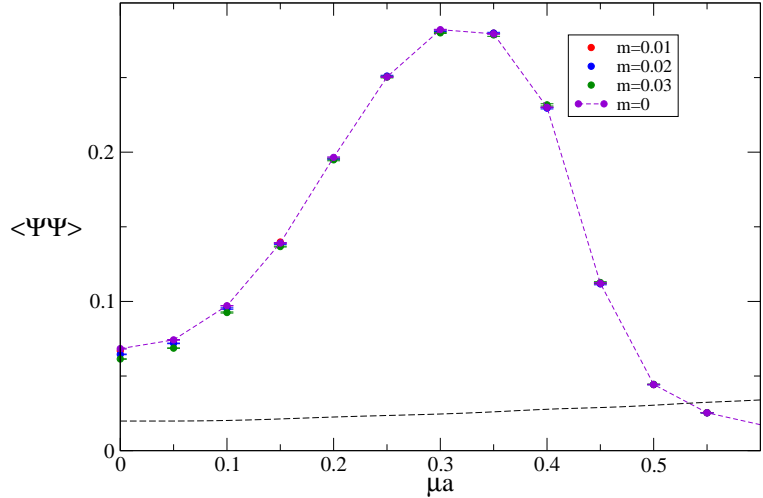}
    %\vspace{-6mm}
    \caption{(Color online)
$\langle  \Psi \Psi \rangle$ vs $\mu$ on $32^3$ for $ja=0.02$ and $ma=0, 0.01, 0.02, 0.03$.
The dashed line shows the same quantity evaluated for $m=0$ for free fields.}
\label{fig:muscan_exciton_varym}
\end{figure}
\begin{figure}[htb]
    \centering
    \includegraphics[width=10.0cm]{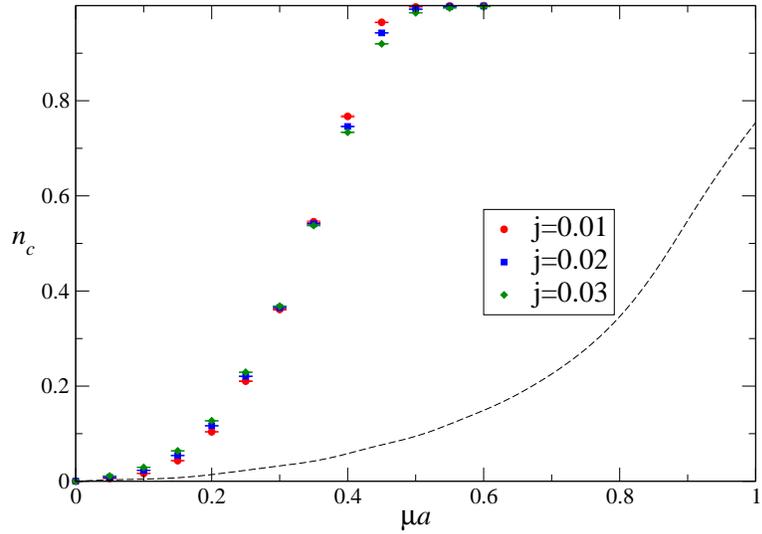}
    %\vspace{-6mm}
    \caption{(Color online)
Carrier density $n_c$ vs $\mu$ on $32^3$, $m=0$ and $j=0.01, 0.02, 0.03$.
The dashed line shows the same quantity evaluated with $j=0.01$ for free fields.
}
\label{fig:carrier_density}
\end{figure}

Figure~\ref{fig:muscan_exciton} shows the exciton condensate $\langle\Psi\Psi\rangle$ as a function of
$\mu$  for three different $j$.  The figure shows the same broad
features as Fig.~\ref{fig:overview}, namely a rapid rise to a fairly sharp
maximum at $\mu a\approx0.3$, followed by a still more rapid fall;  the signal
is very small indeed by $\mu a=0.6$. As we shall see in
Fig.~\ref{fig:carrier_density}, at this value of $\mu$
the system has reached saturation with a maximum possible density of
particle-hole pairs consistent with the Pauli exclusion principle on a fixed
lattice; our model can only be interpreted as a description of bilayer graphene
for values of $\mu$ much smaller than this.

The dashed lines in Fig.~\ref{fig:muscan_exciton} show $\langle\Psi\Psi\rangle$
evaluated using the same measurement code but with $g^2$ set to zero, yielding
the value for free fields. Since the (U(1)$\otimes$U(1)$_\varepsilon)^2$
symmetry is manifest for $j=0$ the free-field condensate must vanish in
this limit, and the curves are consistent with this expectation. 
The large
disparity between $\langle\Psi\Psi\rangle_{\rm int}$ and
$\langle\Psi\Psi\rangle_{\rm free}$ notable in the range $0.2\lapprox\mu
a\lapprox0.4$ signals that (U(1)$\otimes$U(1)$_\varepsilon)^2$ is surely
spontaneously broken here.
Close
inspection of the figure reveals that $\langle\Psi\Psi\rangle_{\rm free}$ rises
monotonically, but not quite smoothly, with $\mu$ until reaching a maximum at 
$\mu a\lapprox0.9$. The disparity with the apparent saturation observed in the
interacting model will be further discussed below. 
The barely visible wiggles are probably a finite volume
artifact similar to that noted in studies of another system with a Fermi
surface~\cite{Hands:2002mr}. 
Figure~\ref{fig:muscan_exciton_varym} plots the same scan but this time showing that
the effect of varying $m$ is negligible except for the very smallest values of
$\mu$. Since the operator $\Psi\Psi$ is constructed to
be conjugate to $j$, not $m$, this is as expected.

\begin{figure}[htb]
    \centering
    \includegraphics[width=10.0cm]{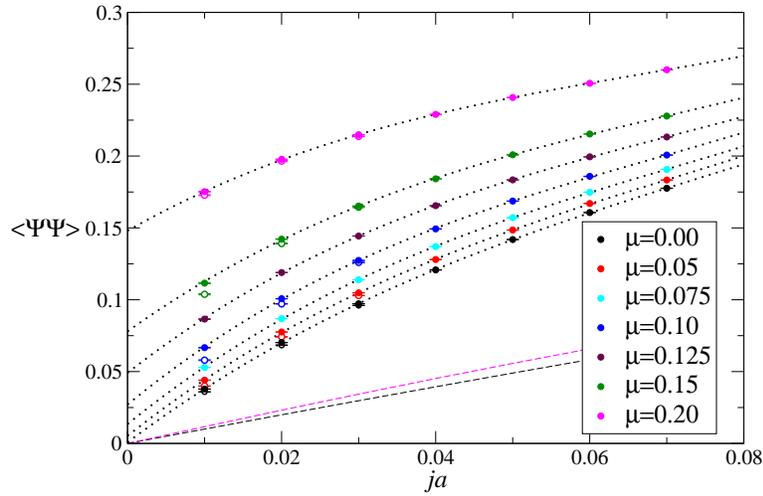}
    %\vspace{-6mm}
    \caption{(Color online)
$\langle  \Psi\Psi \rangle$ vs $j$ for $m=0$ and various $\mu$ on
$32^3$ (open) and $48^3$ (closed symbols). Dotted lines show fits to
eqn.~(\ref{eq:quadrat}).
Dashed lines show the same quantities evaluated for $\mu=0,0.2$ on $48^3$ for
free fields.}
\label{fig:jto0}
\end{figure}
In order to interpret the condensate data it is necessary to
extrapolate $j\to0$. Figure~\ref{fig:jto0} shows $\langle\Psi\Psi\rangle$ for
several $j$ on two different volumes, together with extrapolations of the
form 
\begin{equation}
\langle\Psi\Psi\rangle=\langle\Psi\Psi(j=0)\rangle+Aj+Bj^2+Cj^3.
\label{eq:quadrat}
\end{equation}
Taking finite volume effects into account, it seems that at least for $\mu a\geq0.10$ the
fitted intercept is non-vanishing, confirming the spontaneous breaking of
particle-hole symmetry due to excitonic condensation
$\langle\Psi\Psi\rangle\not=0$. 
\begin{figure}[htb]
    \centering
    \includegraphics[width=10.0cm]{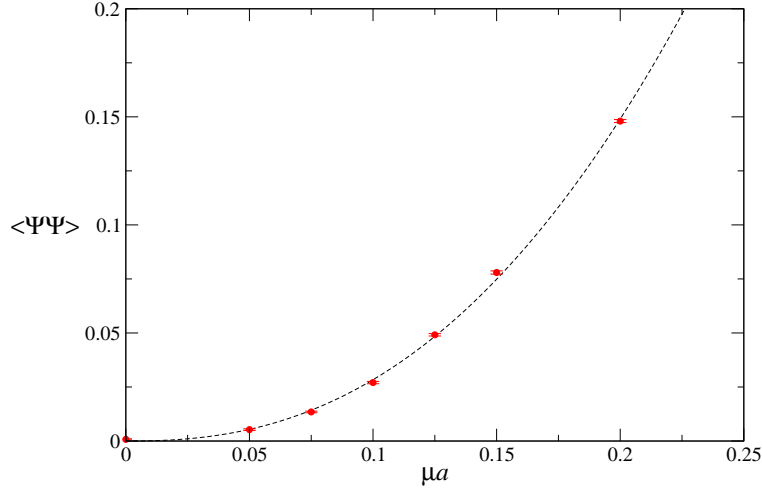}
    %\vspace{-6mm}
    \caption{(Color online)
$\langle\Psi\Psi(j=0)\rangle$ vs $\mu$ on $48^3$ 
fitted to a power law for $\mu=0.05 - 0.20$. 
The dashed line corresponds to exponent $a_2=2.39(2)$.
}
\label{fig:excitonvsmu}
\end{figure}
The extrapolated condensate 
is shown fitted to a power law of the form
$\langle\Psi\Psi(j=0)\rangle=a_1\mu^{a_2}$ in Fig.~\ref{fig:excitonvsmu}: the
fitted parameters are
\begin{equation}
a_1=7.0(2);\;\;\;\; a_2=2.39(2).
\label{eq:fit1}
\end{equation}
The power-law rise is more rapid than would be expected from a
BCS-style
mechanism driven by condensation of particle-hole pairs in the immediate
vicinity of a Fermi surface. This is because in a BCS condensation 
the density of available pairing
states scales with the area of the Fermi surface, $\propto\mu^{d-1}$ in $d$
space dimensions. Despite this somewhat empirical approach, the non-linear
increase of $\langle\Psi\Psi\rangle$ with $\mu$ is a robust conclusion at
variance with a conventional weakly-interacting BCS scenario.

Next we consider the carrier density $n_c$ defined in (\ref{eq:cdensity}), and shown
in Fig.~\ref{fig:carrier_density}. This rises monotonically from zero with $\mu$ until
$\mu a\sim0.5$, when saturation sets in; the effect of $j\not=0$ is to round off
this behaviour by reducing the carrier susceptibility $\vert\partial
n_c/\partial\mu\vert$ slightly. Once again, the contrast with the free-field
behaviour, which only reaches saturation at $\mu a \approx1.3$ and is
shown by the dashed line, is marked.

How should we interpret the finding that $n_c^{\rm int}\gg n_c^{\rm
free}$? 
For degenerate fermions the carrier density, remembering to count both
particle and hole states, is given by  $n_c=k_F^2/2\pi$. 
For free massless fermions the Fermi energy $\mu$ is 
equal to Fermi momentum $k_F$; 
If we wish to
retain the notion of a Fermi surface (albeit one distorted by exciton
condensation) for the interacting system, 
we are forced to conclude $\mu\approx
E_F<k_F$ implying strong self-binding, i.e. the degenerate fermions have a large 
negative contribution to their bulk energy. This is a feature of working near a
QCP, and was not observed, e.g. in studies of relatively weakly-interacting
systems at non-zero density such as the Gross-Neveu model in
2+1$d$~\cite{Hands:1992ck} where interactions are suppressed by $1/N_f$, or two
color QCD~\cite{Cotter:2012mb} where the quark density $n_q\gapprox n_q^{\rm
free}$ all the way to saturation.

\begin{figure}[htb]
    \centering
    \includegraphics[width=10.0cm]{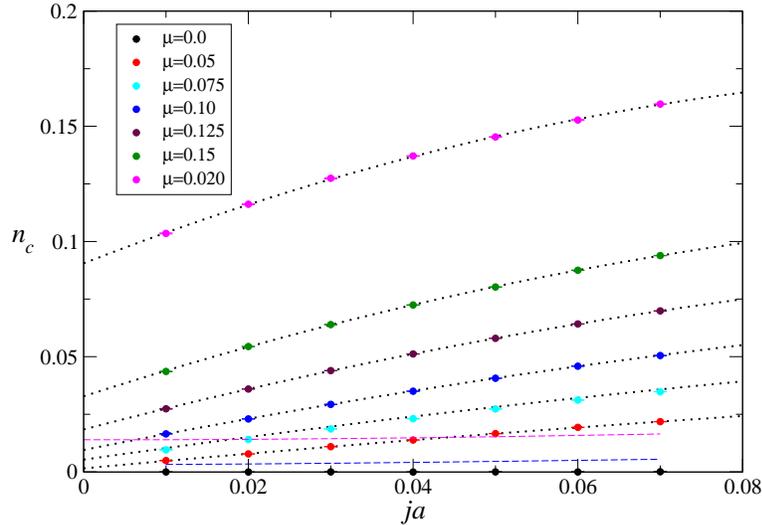}
     %vspace{-6mm}
    \caption{(Color online)
Carrier density $n_c$ vs $j$ on $48^3$ for various $\mu$. Dotted lines show a
quadratic extrapolation $j\to0$. Dashed lines show
the same quantity evaluated for free fields with $\mu=0.1,0.2$.
}
\label{fig:dndj}
\end{figure}
\begin{figure}[!htb]
    \centering
    \includegraphics[width=10.0cm]{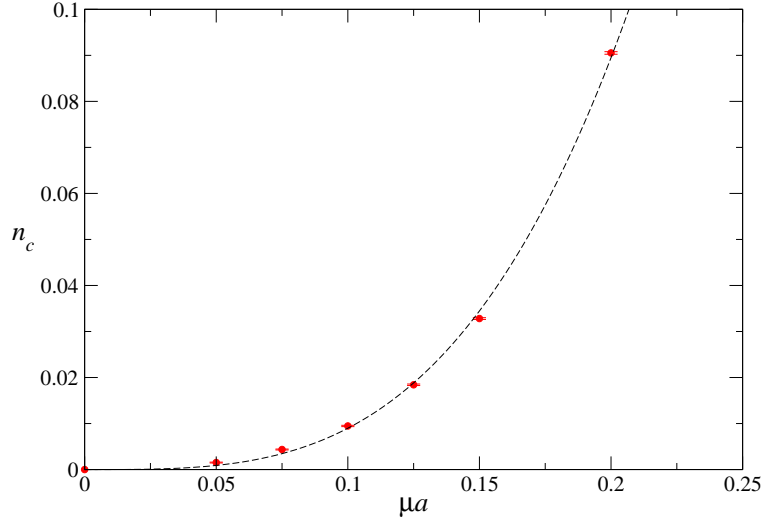}
     %\vspace{-6mm}
    \caption{(Color online)
Carrier density $n_c(j=0)$ vs $\mu$ on $48^3$ 
fitted to a power law for $\mu=0.05 - 0.20$. 
The dashed line corresponds to exponent $b_2=3.32(1)$.
}
\label{fig:cdensityvsmu}
\end{figure}
As before, the region of physical interest is for
$\mu$ well below saturation: Figure~\ref{fig:dndj} plots the variation of $n_c$
with source strength $j$, togther with a quadratic extrapolation to $j=0$, 
showing that the effect of $j\not=0$ for this observable is regular but 
certainly not negligible.
Finally Fig.~\ref{fig:cdensityvsmu} plots $n_c(\mu;j=0)$ together with a power
law fit $n_c=b_1\mu^{b_2}$. The fitted parameters are 
\begin{equation}
b_1=18.6(4);\;\;\;\;\; b_2=3.32(1).
\label{eq:fit2}
\end{equation}
As expected, the fitted value of $b_2$ considerably exceeds
the naive expectation $n_c\propto\mu^d$ based on a weakly-interacting
system.

\begin{figure}[!htb]
    \centering
    \includegraphics[width=10.0cm]{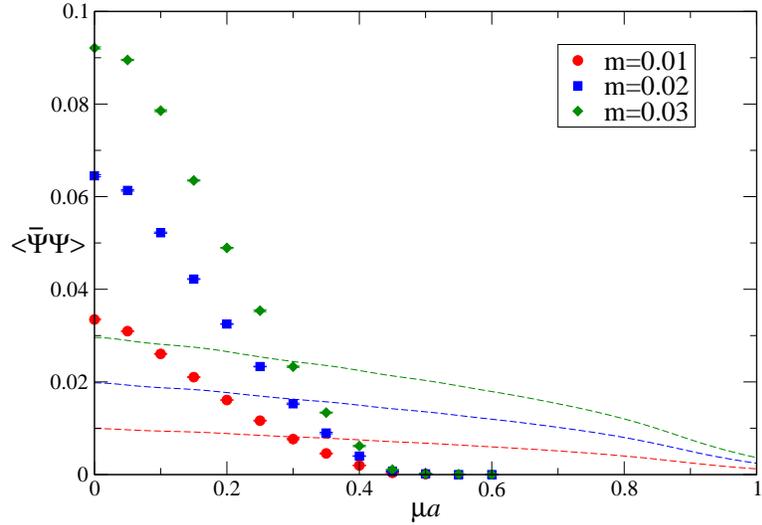}
    %\vspace{-6mm}
    \caption{(Color online)
Chiral condensate $\langle\bar\Psi\Psi\rangle$ 
vs $\mu$ on $32^3$ for $j=0.02$ and $m=0.01, 0.02, 0.03$. Dashed lines show the
same quantity evaluated for free fields.
}
\label{fig:chiral}
\end{figure}
In Fig.~\ref{fig:chiral} we show the chiral condensate order parameter
$\langle\bar\Psi\Psi\rangle$ as a function of $\mu$ for various values of the
symmetry breaking parameter $m$ at fixed $ja=0.02$. Its magnitude at $\mu=0$
falls approximately linearly with $m$ implying restoration of chiral symmetry as
$m\to0$. Even so, it exceeds the free-field value by over a factor of two,
reflecting the vicinity of the QCP.  
Note though that $\vert\langle\bar\Psi\Psi\rangle\vert$, a measure of
the density of particle-hole pairs in the condensate, is roughly one-third of 
the peak value of the exciton condensate $\vert\langle\Psi\Psi\rangle\vert$ seen in
Fig.~\ref{fig:muscan_exciton}.
As $\mu$ increases $\vert\langle\bar\Psi\Psi\rangle\vert$
falls monotonically reflecting the increasing
difficulty of particle-hole pairing {\em within} a layer as the biassing voltage
rises. We deduce that near the QCP the impact of the biassing voltage is to
favour inter-layer over intra-layer pairing; indeed the inter-layer pairing is
suppressed completely, falling below even the free-field value,
 by $\mu a =0.5$ where saturation sets in. 

%{\it
\begin{figure}[htb]
    \centering
    \includegraphics[width=10.0cm]{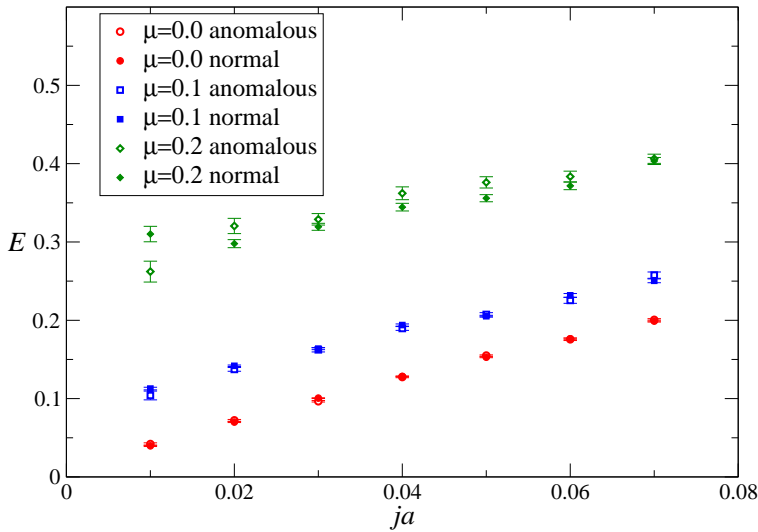}
    %\vspace{-6mm}
    \caption{(Color online)
Normal and anomalous fermion energies $E(k=0)$  vs $j$ on $48^3$ for $\mu=0, 0.1, 0.2$.
\label{fig:quasispectrum}
}
\end{figure}
Finally we report on a preliminary calculation of the spectrum of quasiparticle
excitations, obtained from analysis of the following fermion correlators:
\begin{eqnarray}
C_N(\vec k,t)&=&\sum_{\vec x}\langle\psi(\vec 0,0)\bar\psi(\vec x,t)\rangle
e^{-i\vec k.\vec x};\nonumber\\
C_A(\vec k,t)&=&\sum_{\vec x}\langle\psi(\vec 0,0)\bar\phi(\vec x,t)\rangle
e^{-i\vec k.\vec x}.
\end{eqnarray}
Due to the form of the staggered fermion action (\ref{eq:slatt}) 
the set of two-dimensional sites $\vec x$ only includes 
those displaced from the origin by an even number of lattice spacings in any
direction, and the physically accessible momenta have $k_i=2\pi n_i/L_s$ with
$n_i=0,1,\ldots,L_s/4$. We distinguish between the {\em normal\/} propagator
$C_N$
describing carrier motion within a layer, and {\em anomalous\/} propagator $C_A$
describing inter-layer hopping, which relates eg. destruction of an electron on
layer 1 to creation of an anti-hole on layer 2. On a finite system $C_A$ is
non-vanishing only for $j\not=0$. 

In this first study we have considered $\vec k=\vec0$ only.
In accordance with a study of
quasi-particle propagation in a thin-film BCS superfluid~\cite{Hands:2001aq} we find
that the correlator signal resides in ${\rm Re}(C_N)$ and ${\rm Im}(C_A)$, and that in the
chiral limit $m\to0$ $C_N(t)\equiv0$ for $t$ even and $C_A(t)\equiv0$ for $t$
odd. We thus fit the correlators on every second timeslice for the excitation energy $E$ using the forms
\begin{eqnarray}
{\rm Re}(C_N(\vec k,t))&=&Ae^{-Et}+Be^{-E(L_t-t)};\nonumber\\
{\rm Im}(C_A(\vec k,t))&=&C(e^{-Et}-e^{-E(L_t-t)}),
\end{eqnarray}
where in general $A\not=B$ for $\mu\not=0$. The resulting energies are shown for small
$\mu$ as a function of 
$j$ in Fig.~\ref{fig:quasispectrum}. Two features are apparent: firstly normal
and anomalous channels yield consistent results, as
expected~\cite{Hands:2001aq}, although the normal data have smaller errorbars;
secondly the extrapolation $j\to0$ appears smooth, and suggests
$\lim_{j\to0}E(j)>0$ for $\mu\not=0$. In other words, a voltage bias induces
anomalous propagation indicative of particle-hole mixing, a
manifestation of excitonic condensation $\langle\Psi\Psi\rangle\not=0$.
This should also result in a non-vanishing energy gap at the Fermi surface, but
confirmation requires a study of propagation with $\vec
k\not=\vec0$~\cite{Hands:2002mr}.
%}

\section{Discussion}
\label{sec:discussion}

In this paper we have set out an effective (albeit simplified) field theory for low-energy
charge transport in voltage-biased bilayer graphene, and shown how it can be
simulated using orthodox lattice field theory methods, because its action in
Euclidean metric is real.  An interesting feature of the numerical formulation
is that it is possible to run in the chiral limit $m=0$ so long as the
$\psi\phi$ coupling $j\not=0$.  There are formal similarities to QCD with
non-zero isopsin density~\cite{Kogut:2002zg}; however the resulting dynamics
differs sharply. While QCD is an asymptotically-free theory implying that
eventually a
weakly-coupled description becomes valid as $\mu\to\infty$, here the field
correlations remain strong at all scales, even in the absence of confinement,
due to the vicinity of the QCP.  For this reason the model is of intrinsic
theoretical interest independent of any possible physical applications
for graphene.

Precise location of the QCP by numerical means has proved challenging;
nevertheless the curvature of the chiral condensate data
$\langle\bar\Psi\Psi(m)\rangle$ of Fig.~\ref{fig:chirallimit} are suggestive of
a critical scaling $\langle\bar\Psi\Psi\rangle\propto m^{1\over\delta}$ expected
at or near a QCP. Equation
of state fits predict $\delta$ in the range 2.7 (the value for monolayer
graphene with $N_f=2$)~\cite{AHS1} to  5.5 (the value in the strong coupling
limit $N_f=N_{fc}\approx4.8(2)$)~\cite{Hands:2008id}. Considerably more work
would be needed to confirm this quantitatively.

Our main results in this first study are therefore qualitative. 
Runs with $j\not=0$ yield measurements of the exciton condensate
$\langle\Psi\Psi\rangle$ which show a rapid rise as $\mu$ is increased from
zero. The data extrapolated to $j\to0$  suggest that condensate remains
non-vanishing in this limit consistent with spontaneous symmetry breaking and
superfluidity; indeed the data of Fig.~\ref{fig:excitonvsmu} permit a power-law
fit $\langle\Psi\Psi\rangle\propto\mu^{2.4}$. This is notable because a
weak-coupling BCS description of superfluidity predicts the condensate should
scale with the area of the Fermi surface, namely
$\langle\Psi\Psi\rangle\propto\mu$. Similarly, the carrier density
$n_c\propto\mu^{3.3}$ (Fig.~\ref{fig:dndj}), to be contrasted with the
weak-coupling behaviour $n_c\propto\mu^2$. With the resolution we are
working with there is no sign of an {\em onset\/} value of the chemical potential
$\mu_o>0$, such that $n_c=0$, $\langle\Psi\Psi\rangle=0$ for $\mu<\mu_o$. This
is another important contrast with the systems studied in
\cite{Hands:2003dh,Hands:2001aq,Hands:2002mr,Cotter:2012mb,Maas:2012wr,Kogut:2002zg}.
The likely reason is that at the couplings studied there is no mechanism for spontaneous mass generation,
so that the lightest degree of freedom carrying a conserved charge remains
massless. The final interesting observation, shown in Fig.~\ref{fig:chiral}, is
that the chiral conndensate $\langle\bar\Psi\Psi\rangle$ is strongly suppressed
as $\mu$ rises, presumably because of the rapidly-increasing energy cost of a
particle-hole pair {\it within\/} a layer, and is consistent with zero
post-saturation.

Another observation to note is that below saturation both
$n_c\gg n_c^{\rm free}$ and $\langle\Psi\Psi\rangle\gg\langle\Psi\Psi\rangle^{\rm
free}$. Once again, this is indicative of strong correlations among the fields,
such that $E_F<k_F$,
as is the precocious value of $\mu a$ at which saturation sets in. It suggests
that the self-consistent diagrammatic approach of \cite{Kharitonov} (which 
employs large-$N_f$ methods so does
not probe the QCP) may yield an unduly small estimate of the condensate. It must
be stressed, however, that in the absence of a physical scale setting
any phenomenogical applications of the model to real graphene are premature.

\begin{figure}[htb]
    \centering
    \includegraphics[width=10.0cm]{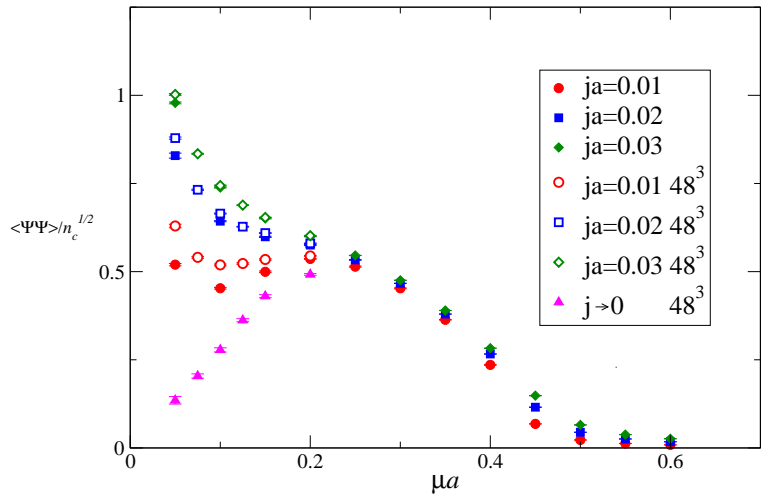}
    %\vspace{-6mm}
    \caption{(Color online)
The ratio $\langle\Psi\Psi\rangle/n_c^{1\over2}$  vs $\mu a $ on $32^3$ (filled)
and $48^3$ (open) for
$ja=0.01,0.02,0.03$, together with the $j\to0$ extrapolation on $48^3$. 
\label{fig:ratio}
}
\end{figure}
In conclusion we claim to have initiated a lattice Monte Carlo study of strongly interacting 
degenerate fermions, which displays significant qualitative differences to
other degenerate systems studied previously. A final question worth discussing is
to what extent the concept of a Fermi surface, either sharp or distorted by
particle-hole excitonic condensation, remains intact in a strongly-interacting
environment. Departures from the canonical weak-coupling are manifested as
anomalous scaling with Fermi energy $\mu$ (see eqns.(\ref{eq:fit1},\ref{eq:fit2})); 
however recall that in 
an interacting Fermi liquid the
relation between particle density and Fermi momentum $k_F$, namely $n_c\propto
k_F^d$, should remain inviolate. In the BCS picture, the density of
condensed particle-hole pairs $\langle\Psi\Psi\rangle$ arising from plane wave
states
within a shell of thickness $\Delta$ around the Fermi surface
implies
\begin{equation}
\langle\Psi\Psi\rangle
\propto \Delta k_F^{d-1} 
\propto \Delta n_c^{{d-1}\over d}.
\label{eq:BCSsc}
\end{equation}
To test whether the scaling (\ref{eq:BCSsc})  is retained even
at strong coupling, 
Fig.~\ref{fig:ratio} plots the ratio $\langle\Psi\Psi\rangle/n_c^{1\over2}$ vs. $\mu$ for
various $j$ on two volumes, together with the $j\to0$ extrapolation on $48^3$. 
It looks plausible
for $\mu a\lapprox0.2$, on the assumption that the gap $\Delta$ has a strong
$\mu$-dependence, which should be the case for near-conformal dynamics. 
It may well prove possible, therefore, to
define a Fermi surface in the vicinity of a quantum critical point.

\section{Acknowledgements}

CPUs used were either Intel(R) Xeon(R) E5420, X5650 or E5-2660. We estimate
that over 2 million core hours of computing time were needed to complete this
project. The authors wish to thank Diamond Light Source, the Oxford Super
Computing Centre (OSC) and the e-Infrastructure South consortium for kindly
allowing them to use extensive computing resources, and specifically Tina
Friedrich at Diamond Light Source, Luke Raimbach, Lino Garcia Tarres and Steven
Young at OSC for help in configuring and maintaining these resources.

\end{document}